\documentclass[Times,4pt,aps,pra,onecolumn,showpacs,amsmath,amssymb,floatfix,footinbib,superscriptaddress,times]{revtex4-1}
\usepackage{graphics,graphicx,epsfig,ulem,epstopdf,bm,longtable,url,datetime}
\usepackage[colorlinks=true,linkcolor=red,urlcolor=red,citecolor=red]{hyperref}
\usepackage{amsmath,amssymb,latexsym,float,amsfonts,subfigure,hyperref,color}
\usepackage[ansinew]{inputenc}
\usepackage[usenames,dvipsnames]{pstricks}
\usepackage[mathlines]{lineno}
\def\footnoterule{\kern -1mm \hrule width 5.8cm \kern 2.2mm}%
\usepackage{tikz,xcolor,hyperref}
\definecolor{lime}{HTML}{A6CE39}
\DeclareRobustCommand{\orcidicon}{%
    \begin{tikzpicture}
    \draw[lime, fill=lime] (0,0)
    circle [radius=0.16]
    node[white] {{\fontfamily{qag}\selectfont \tiny ID}};\draw[white, fill=white] (-0.0625,0.095)
    circle [radius=0.007];
    \end{tikzpicture}
    \hspace{-2mm}}
\foreach \x in {A, ..., Z}
{\expandafter\xdef\csname orcid\x\endcsname{\noexpand\href{https://orcid.org/\csname orcidauthor\x\endcsname}{\noexpand\orcidicon}}}

\begin{document}
\title{Electromagnetic chirality-induced negative refraction with the same amplitude and anti-phase of the two chirality coefficients}

\author{Shun-Cai Zhao\orcidA{}}%
\email[Corresponding author: ]{zscnum1@126.com.}
\affiliation{School of Materials Science and Engineering, Nanchang University, Nanchang 330031, China}
\affiliation{Engineering Research Center for Nanotechnology,Nanchang University, Nanchang 330047,China}
\affiliation{Institute of Modern Physics, Nanchang University, Nanchang 330031,China}

\author{Zheng-Dong Liu}
\email[Co-first author.]{lzdgroup@ncu.edu.cn}
\affiliation{School of Materials Science and Engineering, Nanchang University, Nanchang 330031, China}
\affiliation{Engineering Research Center for Nanotechnology,Nanchang University, Nanchang 330047,China}
\affiliation{Institute of Modern Physics, Nanchang University, Nanchang 330031,China}

\author{Jun Zheng}
\affiliation{School of Materials Science and Engineering, Nanchang University, Nanchang 330031, China}
\affiliation{Engineering Research Center for Nanotechnology,Nanchang University, Nanchang 330047,China}
\affiliation{Institute of Modern Physics, Nanchang University, Nanchang 330031,China}

\author{Gen Li}
\affiliation{School of Materials Science and Engineering, Nanchang University, Nanchang 330031, China}
\affiliation{Engineering Research Center for Nanotechnology,Nanchang University, Nanchang 330047,China}
\affiliation{Institute of Modern Physics, Nanchang University, Nanchang 330031,China}
\begin{abstract}
We suggest a scheme of electromagnetic chirality-induced negative
refraction utilizing magneto-electric cross coupling in a four-level
atomic system.The negative refraction can be achieved with the two
chirality coefficients having the same amplitude but the opposite
phase,and without requiring the simultaneous presence of an
electric-dipole and a magnetic-dipole transition near the same
transition frequency.The simultaneously negative electric
permittivity and magnetic permeability does not require,either.

\end{abstract}

\maketitle
\section{Introduction}
Negative refraction of light,first predicted to occur in materials
with simultaneous negative permittivity and permeability in 1968[1],
has attracted considerable attention in the last decade[2-7].
Materials with negative refraction index promise many surprising and
even counterintuitive electromagnetical and optical effects,such as
the reversals of both Doppler shift and Cerenkov
radiation[4],amplification of evanescent waves[8],subwavelength
focusing[8-10] and so on[11,12].Up to now,there have been several
approaches to the realization of negative refractive index
materials, including artificial composite metamaterials[13,14],
photonic crystal structures[15],transmission line simulation[11]and
chiral media[17-18]as well as photonic resonant materials(coherent
atomic vapor)[19-21].The early proposals for negative refraction
required media with both negative permittivity and permeability
($\varepsilon,\mu$$<$0) in the frequency range of
interest[1,4].However,the typical transition magnetic dipole moments
are smaller than transition electric dipole moments by a factor of
the order of the fine structure constant
($\alpha\approx\frac{1}{137}$).So,it is difficult to obtain a
negative permeability since the smaller typical transition
magnetic-dipole moments.Thus,it's difficult to achieve negative
refraction in the optical region of the spectrum because of the
weakness of the magnetic response.To alleviate this problem,
recently, a chiral route to negative refraction has been suggested
[17,22-24]without simultaneously negative electric permittivity and
magnetic permeability. The key idea is to use a magnetoelectric
cross coupling where the medium's electric polarization is coupled
to the magnetic field of the wave and the medium's magnetization is
coupled to the electric field.Under such conditions,a negative index
of refraction can be achieved without requiring a negative
permeability.The medium in which the electric polarization
\textbf{P} is coupled to the magnetic field \textbf{H} of an
electromagnetic wave and the magnetization \textbf{M} is coupled to
the electric field \textbf{E}:
\begin{equation}
\textbf{p}=\varepsilon_{0}\chi_{e}\textbf{E}+\frac{\xi_{EH}}{c}\textbf{H}\\
\textbf{M}=\frac{\xi_{HE}}{c \mu_{0}}\textbf{E}+\chi_{m}\textbf{H}
\end{equation}
Here $\chi_{e}$ and $\chi_{m}$, and $\xi_{EH}$ and $\xi_{HE}$ are
the electric and magnetic susceptibilities,and the complex chirality
coefficients, respectively.They lead to additional contributions to
the refractive index for one circular polarization[17,23-24]:
\begin{equation}
n=\sqrt{\varepsilon\mu-\frac{(\xi_{EH}+\xi_{HE})^{2}}{4}}+\frac{i}{2}(\xi_{EH}-\xi_{HE})
\end{equation}
One typically chooses the phase such that the chirality coefficients
are imaginary,if $\xi_{EH}=-\xi_{HE}=i\xi$,and Eq.(2) reads
n=$\sqrt{\epsilon\mu}-\xi$.When $\sqrt{\varepsilon\mu}$ is less than
$\xi$($\xi>0$),the negative refraction is obtained without requiring
both $\varepsilon<0$ and $\mu < 0$.In this paper, we propose an
alternative scheme to realized negative refractive.we demonstrate
that the electromagnetic chirality induced negative refraction can
be realized when the two chirality coefficients have the same
amplitude but the opposite phase,which requires the chirality
coefficients not only to be imaginary.And all of these current
suggestions for negative refraction in left-handedness[25-26]
require a strong magnetic dipole transition and a strong
electric-dipole transition near the same transition frequency. This
requirement puts a stringent constraint on the energy level
structure of systems where negative refraction can be achieved. In
this article,the electric-dipole and magnetic-dipole transition will
occur at different transition frequency.

The paper is organized as follows: Section 2 establishes the model,
and presents expressions for the chirality coefficients and
refractive index. Section 3 is devoted to present the numerical
results and to discuss numerical results. Finally, Sec.4 summarizes
the conclusions.

\section{Theoretical model and chirality coefficients}

\begin{figure}[htp]
\center
\includegraphics[width=0.40\columnwidth]{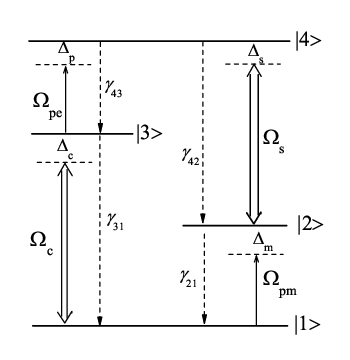}
\caption{Schematic diagram of a four-level atomic system interacting
with the control $\Omega_{c}$,signal $\Omega_{s}$ fields and the
probe field which electric and magnetic components are coupled to
the level pairs$|4\rangle-|3\rangle$ and
$|2\rangle-|1\rangle$,respectively.}
\label{Fig.1}
\end{figure}

We proceed with a detailed description of electromagnetically
induced chiral negative refraction.The four-level configuration of
atoms for consideration is shown in Fig.1.The parity properties of
the atomic states are as follows: levels $|1\rangle$,$|2\rangle$,and
$|3\rangle$ have same parity, which is opposite to the parity of
level $|4\rangle$.Since the two lower levels $|1\rangle$ and
$|2\rangle$ have same parity,the coherent coupling $\Omega_{pm}$
cannot be electric-dipole, but instead can be achieved through the
magnetic-field,and so $\langle2|$$\hat{\vec{\mu}}$$|1\rangle$$\neq0$
where $\hat{\vec{\mu}}$ is the magnetic-dipole operator.The two
upper levels,$|3\rangle$ and $|4\rangle$ have the opposite
parity,the coherent coupling $\Omega_{pe}$ should be electric-dipole
with $\langle4|$$\hat{\vec{d}}$$|3\rangle$$\neq0$ where
$\hat{\vec{d}}$ is the electric dipole operator.As showed in Figure
1,three electromagnetic fields are introduced to couple the four
states:The electric(\textbf{E})and magnetic(\textbf{B}) components
of the probe light(corresponding Rabi frequency
$\Omega_{pe}$$=\frac{\vec{E_{P}}\cdot\vec{d_{34}}}{\hbar}$
,$\Omega_{pm}$$=\frac{\vec{B_{P}}\cdot\vec{\mu_{12}}}{\hbar}$)interact
with the transitions $|3\rangle$ and$|4\rangle$ as well
as$|2\rangle$ and$|1\rangle$,respectively.Hence, the electric and
magnetic components of the probe field with the same frequency
$\omega_{p}$ drive the two transitions $|3\rangle$ -$|4\rangle$ and
$|2\rangle$-$|1\rangle$,simultaneously. The control field with Rabi
frequency denoted by $\Omega_{c}$ pumps atoms in level $|1\rangle$
into upper level $|3\rangle$.According to parity selection rules,
the transition $|1\rangle$-$|3\rangle$ is assumed to be a two-photon
process as stated in [19].The strong signal field couples levels
$|2\rangle$ and $|4\rangle$ with Rabi frequency $\Omega_{s}$.In the
rotating-wave and dipole approximations, the Hamiltonian of the
system can be read in the form
\begin{eqnarray}
H=\sum_{i=1}^{4}\hbar\omega_{i}|i\rangle\langle
i|-\hbar(\Omega_{pm}e^{-i(\omega_{p}t+\theta_{pm})}|2\rangle\langle1|+\Omega_{c}e^{-i(2\omega_{c}t+\theta_{c})}|3\rangle\langle1|
\nonumber\\+\Omega_{pe}e^{-i(\omega_{p}t+\theta_{pe})}|4\rangle\langle3|+\Omega_{s}e^{-i(\omega_{s}t+\theta_{s})}|4\rangle\langle2|+c.c.)
\end{eqnarray}
Where $\omega_{i}(i=c,s)$ are the frequencies of the control and
signal fields,respectively.And $\theta_{i}(i=pe,pm,c,s)$ represent
phases of the electric and magnetic components of the probe field,
control and the signal fields, respectively.When the probe field is
weak,i.e.$\Omega_{pe}$,$\Omega_{pm}$
$\ll$$\Omega_{c}$$<$$\Omega_{s}$.We find the first-order
perturbation solution of the Liouville equation in the steady-state
\begin{equation}
\rho_{43}=\alpha_{EE}\emph{\textbf{E}}+\alpha_{EH}\emph{\textbf{B}},\\
\rho_{21}=\alpha_{HE}\emph{\textbf{E}}+\alpha_{HH}\emph{\textbf{B}}
\end{equation}
where the coefficients $\alpha_{EE}$,$\alpha_{EH}$,$\alpha_{HH}$ and
$\alpha_{HE}$ are given by
\begin{eqnarray}
\alpha_{EE}=\frac{A_{0}\Omega_{c}^{2}d_{34}(A_{11}A_{12}+A_{13})}{D_{0}D_{1}+D_{2}\Omega_{s}^{2}+\Omega_{s}^{4}}
\end{eqnarray}
\begin{eqnarray}
\alpha_{EH}=\frac{e^{i\theta}A_{0}\mu_{12}\Omega_{c}\Omega_{s}[A_{21}-(\Gamma_{2}+i\Delta_{c})(A_{22}-A_{23}\Omega_{c}^{2}-\gamma_{31}\Omega_{s}^{2})]}{D_{0}D_{1}+D_{2}\Omega_{s}^{2}+\Omega_{s}^{4}}
\end{eqnarray}
\begin{eqnarray}
\alpha_{HE}=\frac{e^{i\theta}A_{0}d_{34}\Omega_{c}\Omega_{s}\{A_{41}(\Gamma_{2}+i\Delta_{c})\Omega_{c}^{2}+(i\Delta_{c}-\Gamma_{2})[\gamma_{31}\Omega_{s}^{2}-A_{42}\Omega_{c}^{2}+(i\Gamma_{6}+\Delta_{c}-\Delta_{p})A_{43}]}{D_{0}D_{1}+D_{2}\Omega_{s}^{2}+\Omega_{s}^{4}}
\end{eqnarray}
\begin{eqnarray}
\\\alpha_{HH}=\frac{A_{0}\mu_{12}A_{31}(i\Delta_{c}-\Gamma_{2})\{A_{33}[(\Gamma_{5}+i\Delta_{p})A_{32}+\Omega_{c}^{2}]+A_{32}\Omega_{s}^{2}\}}{D_{0}D_{1}+D_{2}\Omega_{s}^{2}+\Omega_{s}^{4}}\nonumber
\\-\frac{A_{0}\mu_{12}\Omega_{c}^{2}(\Gamma_{2}+i\Delta_{c})\{(\gamma_{31}-A_{33})[(\Gamma_{5}+i\Delta_{p})A_{32}+\Omega_{c}^{2}]-(A_{32}+\gamma_{31})\Omega_{s}^{2}\}}{D_{0}D_{1}+D_{2}\Omega_{s}^{2}+\Omega_{s}^{4}}
\end{eqnarray}
and
\begin{equation}
A_{0}=\frac{i}{\hbar(\Gamma_{2}^{2}\gamma_{31}+\gamma_{31}\Delta_{c}^{2}+4\Gamma_{2}\Omega_{c}^{2})},\nonumber\\
A_{11}=\gamma_{31}(\Gamma_{2}-i\Delta_{c})+2\Gamma_{1}[\Gamma_{3}+i(\Delta_{c}+\Delta_{p})]
\end{equation}
\begin{equation}
A_{12}=(\Gamma_{1}+i\Delta_{m})[\Gamma_{6}-i(\Delta_{c}-\Delta_{p})]+\Omega_{c}^{2},\nonumber\\
A_{13}=\Omega_{s}^{2}[i\gamma_{31}\Delta_{c}-\Gamma_{2}(\gamma_{31}-2\Gamma_{6}+2i\Delta_{c}-2i\Delta_{p})]
\end{equation}
\begin{equation}
A_{21}=(\Gamma_{2}-i\Delta_{c})(\Gamma_{3}+\Gamma_{6}+2i\Delta_{p})(\Gamma_{2}\gamma_{31}+i\gamma_{31}\Delta_{c}+\Omega_{c}^{2}),\nonumber\\
A_{23}=\Gamma_{3}-\gamma_{31}+\Gamma_{6}+2i\Delta_{p}
\end{equation}
\begin{equation}
A_{22}=\gamma_{31}(\Gamma_{1}+i\Delta_{m})[-\Gamma_{3}-i(\Delta_{c}+\Delta_{p})],\nonumber\\
A_{31}=-\Gamma_{2}\gamma_{31}-i\gamma_{31}\Delta_{c}-\Omega_{c}^{2}
\end{equation}
\begin{equation}
A_{32}=\Gamma_{3}+i(\Delta_{c}+\Delta_{p}),A_{33}=\Gamma_{6}+i(\Delta_{p}-\Delta_{c}),A_{41}=\Gamma_{3}+\Gamma_{6}+2i\Delta_{p},\nonumber\\
\end{equation}
\begin{equation}
A_{42}=\Gamma_{3}+\gamma_{31}+i(\Delta_{c}+\Delta_{p}),A_{43}=\gamma_{31}(\Delta_{p}-i\Gamma_{5})+i\Omega_{c}^{2},\nonumber\\
\end{equation}
\begin{equation}
D_{0}=(\Gamma_{1}+i\Delta_{m})[\Gamma_{6}-i(\Delta_{c}-\Delta_{p})]+\Omega_{c}^{2},
D_{1}=(\Gamma_{5}+i\Delta_{p})[\Gamma_{3}+i(\Delta_{c}+\Delta_{p})]+\Omega_{c}^{2},\nonumber\\
\end{equation}
\begin{equation}
D_{2}=(i\Gamma_{6}+\Delta_{c}-\Delta_{p})(\Delta_{p}-i\Gamma_{5})+(\Gamma_{1}+i\Delta_{m})[\Gamma_{3}+i(\Delta_{c}+\Delta_{p})]-2\Omega_{c}^{2}\nonumber\\
\end{equation}
with the coherence damping coefficient given by
$\Gamma_{1}=\frac{1}{2}(\gamma_{1}+\gamma_{21})+\gamma_{c}$,$\Gamma_{2}=\frac{1}{2}(\gamma_{1}+\gamma_{31})+\gamma_{c}$,
$\Gamma_{3}=\frac{1}{2}(\gamma_{1}+\gamma_{42}+\gamma_{43})+\gamma_{c}$,$\Gamma_{4}=\frac{1}{2}(\gamma_{21}+\gamma_{42}+\gamma_{43})+\gamma_{c}$,
$\Gamma_{5}=\frac{1}{2}(\gamma_{31}+\gamma_{42}+\gamma_{43})+\gamma_{c}$,$\Gamma_{6}=\frac{1}{2}(\gamma_{31}+\gamma_{21})$,
 in which $\gamma_{c}$ denotes the collisional dephasing rate,and $\gamma_{1}=0$.The detunings of the applied fields are respectively defined by
$\Delta_{p}=\omega_{p}-(\omega_{4}-\omega_{3})$,
$\Delta_{c}=2\omega_{c}-(\omega_{3}-\omega_{1})$,$\Delta_{s}=\omega_{s}-(\omega_{4}-\omega_{2})$,
$\Delta_{m}=\omega_{p}-(\omega_{2}-\omega_{1})$ and here we depict
the electric-dipole and magnetic-dipole transitions be different by
seting $\Delta_{m}\neq\Delta_{p}$ (i.e. two transition frequencies
are not near the same frequency in this atomic system)[20].The
relative phase of the signal, control and probe electric fields and
probe magnetic field is
$\theta=\theta_{pe}+\theta_{c}-\theta_{pm}-\theta_{s}$.It is well
known that the phases of the electric and magnetic components of an
electromagnetic field are identical in a nonconductor [27]. Hence,
in our scheme,$\theta_{pe}=\theta_{pm}$ and the relative phase
becomes the phase difference of the control and pump
fields,i.e.,$\theta=\theta_{c}-\theta_{s}$.

The ensemble electric polarization and magnetization of the atomic
medium to the probe field are given by
$\vec{P}$=$N\vec{d_{12}}\rho_{21}$ and
$\vec{M}$=$N\vec{\mu_{34}}\rho_{43}$,respectively,where N is the
density of atoms.Then the coherent cross-coupling between electric
and magnetic dipole transitions driven by the electric and magnetic
components of the probe field may lead to chirality
[17,23-24].Substituting equations (4) into the formula for the
ensemble electric polarization and magnetization, we have the
relations
\begin{equation}
\textbf{P}=a_{1} \textbf{E}+a_{2} \textbf{B},
 \textbf{M}=a_{3}\textbf{E}+a_{4} \textbf{B}
\end{equation}
where
\begin{equation}
a_{1}=Nd_{12}\alpha_{HE},\nonumber\\a_{2}=Nd_{12}\alpha_{HH},
\end{equation}
\begin{equation}
a_{3}=N\mu_{34}\alpha_{EE},\nonumber\\a_{4}=N\mu_{34}\alpha_{EH}.
\end{equation}
Considering both electric and magnetic local field
effects[27,28],\emph{\textbf{E }}and \emph{\textbf{B}} in
equation(10)must be replaced by the local fields
\begin{equation}
\textbf{E}_{L}=\textbf{E}+\frac{\textbf{P}}{3\varepsilon_{0}},\nonumber\\\textbf{B}_{L}=\mu_{0}(\textbf{H}+\frac{\textbf{M}}{3})
\end{equation}
As a result, we obtain
\begin{eqnarray}
\textbf{P}=\frac{3\varepsilon_{0}(\mu_{0}a_{2}a_{1}-\mu_{0}a_{3}a_{2}-3a_{1})}{\mu_{0}a_{3}\alpha_{EB}+3a_{1}-\mu_{0}a_{4}a_{1}-9\varepsilon_{0}+3\mu_{0}\varepsilon_{0}a_{4}}\textbf{E}
\nonumber\\
+\frac{-9\mu_{0}\varepsilon_{0}a_{2}}{\mu_{0}a_{3}a_{2}+3a_{1}-\mu_{0}a_{4}a_{1}-9\varepsilon_{0}+3\mu_{0}\varepsilon_{0}a_{4}}\textbf{H}\label{},
\end{eqnarray}
\begin{eqnarray}
\textbf{M}=\frac{9\varepsilon_{0}a_{3}}{\mu_{0}a_{4}a_{1}+9\varepsilon_{0}-\mu_{0}a_{3}a_{2}-3a_{1}-3\varepsilon_{0}\mu_{0}a_{4}}\textbf{E} \nonumber\\
+\frac{3(\mu_{0}a_{3}a_{2}-\mu_{0}a_{4}a_{1}+3\varepsilon_{0}\mu_{0}a_{4})}{\mu_{0}a_{4}a_{1}+9\varepsilon_{0}-\mu_{0}a_{3}a_{2}-3a_{1}-3\varepsilon_{0}\mu_{0}a_{4}}\textbf{H}
\end{eqnarray}
By comparison with equation (1), we obtain the permittivity and the
permeability, and the complex chirality coefficients:
\begin{eqnarray}
\varepsilon=1+\chi_{e}=\frac{6a_{1}+9\varepsilon_{0}+\mu_{0}[2a_{3}a_{2}-a_{4}(2a_{1}+3\varepsilon_{0})]}{-3a_{1}+\mu_{0}[-a_{3}a_{2}+a_{4}(a_{1}-3\varepsilon_{0})]+9\varepsilon_{0}},
\end{eqnarray}
\begin{eqnarray}
\mu=1+\chi_{m}=\frac{-3\alpha_{EE}+2\mu_{0}[\alpha_{BE}\alpha_{EB}-\alpha_{BB}(\alpha_{EE}-3\varepsilon_{0})]+9\varepsilon_{0}}{-3\alpha_{EE}+\mu_{0}[-a_{3}a_{2}+a_{4}(a_{1}-3\varepsilon_{0})]+9\varepsilon_{0}},
\end{eqnarray}
\begin{eqnarray}
\xi_{EH}=\frac{9c\mu_{0}a_{2}\varepsilon_{0}}{-3a_{1}+\mu_{0}[-a_{3}a_{2}+a_{4}(a_{1}-3\varepsilon_{0})]+9\varepsilon_{0}},
\end{eqnarray}
\begin{eqnarray}
\xi_{HE}=\frac{9c\mu_{0}a_{3}\varepsilon_{0}}{-3a_{1}+\mu_{0}[-a_{3}a_{2}+a_{4}(a_{1}-3\varepsilon_{0})]+9\varepsilon_{0}}
\end{eqnarray}

In the above,we obtained the expressions for the electric
permittivity and magnetic permeability,and the complex chirality
coefficients of the atomic media. Substituting equations
from(13)to(16)into(2),the expression for refractive index can also
be presented.In the section that follows, we will show that the
negative refractive index of the atomic system can be obtained
without requiring simultaneously negative both permittivity and
permeability.

\section{Results and discussion}

Before doing these calculations, we need to fix several key
parameters such as spontaneous emission rate, wavelength and atomic
density in these equations.In the model configuration,the transition
from$|2\rangle$to$|1\rangle$is magnetic dipole allowed and others
are electrical dipole allowed.The typical value of the spontaneous
emission rate of atomic electric dipole transitions is the magnitude
of $10^{8}$ Hz.The spontaneous emission rate of atomic magnetic
dipole transitions is in general smaller than that of atomic
electric dipole transitions by four magnitude.Thus,in our numerical
calculations,the spontaneous emission rates are scaled by
$\gamma=10^{8}s^{-1}$:$\gamma_{21}=\gamma_{43}\times(\frac{1}{137})^{2}$[20],
$\gamma_{31}=\gamma_{42}=\gamma_{c}=1\gamma$ for simplicity.The
typical optical wavelength for the
transitions$|4\rangle$$\rightarrow$$|3\rangle$ and
$|2\rangle$$\rightarrow$$|1\rangle$ is selected to be 600 nm.The
dipole moments $d_{12}$ and $\mu_{34}$ are estimated by the
relation$\sqrt{3\hbar\Gamma_{ij}\lambda^{3}/8\pi^{2}}$.In the
present calculations, we choose the density of atoms N to be
$5\times10^{24}m^{-3}$.Dense vapor is required so that the atomic
density should be larger than $10^{24}m^{-3}$[20].The detuning of
the strong signal field is set as $\Delta_{s}$=0 with Rabi frequency
$\Omega_{s}=20\gamma$.As mentioned previously, The requirement of a
strong magnetic dipole transition and a strong electric-dipole
transition near the same transition frequency puts a stringent
constraint on the energy level structure of systems where negative
refraction can be achieved[20].In our scheme,the electric-dipole and
magnetic-dipole transition occur at different transition
frequency,because the magnetic component of probe field and the
control field detunings are assumed to be the same,
$\Delta_{m}=\Delta_{c}=0.001\gamma$.

\begin{figure}[htp]
\center
\includegraphics[width=0.40\columnwidth]{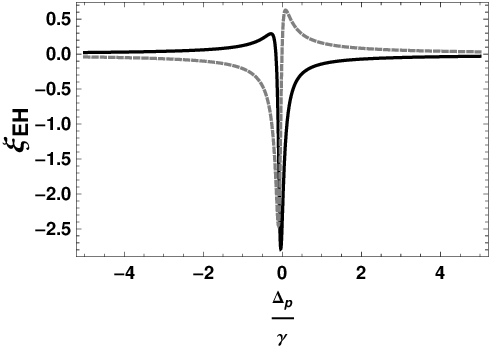}\includegraphics[width=0.40\columnwidth]{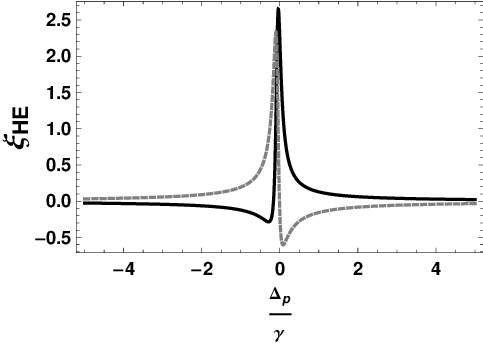}
\caption{Real(solid lines) and imaginary(dashed lines) parts of the
chirality coefficients as a function of the rescaled detuning
parameter $\Delta_{p}/\gamma$ for $\theta = 1/5 \pi$, $\Omega_{c} =
1.3 \gamma$, and the other parameters given in the text.}
\label{Fig.2}
\end{figure}
Fig.2 shows the real and imaginary parts of the chirality
coefficients $\xi_{EH}$ and $\xi_{HE}$ as the function of
$\Delta_{p}/\gamma$.It is observed their real and imaginary parts
have the same order of magnitude but the opposite
phase,simultaneously.And the similar conclusion can also be drawn
when varying the $\Omega_{c}$= with 0.4$ \gamma$, 0.8$\gamma$ ,or
$\theta$ = 3/2$ \pi$,$\Omega_{c}$ =1 $\gamma$,1.4
$\gamma$,1.8$\gamma$,respectively. The sum of the two chirality
coefficients $\xi_{EH}$ and $\xi_{HE}$ will vanish in Eq.(2).And
negative refraction can be possible to obtain when the chirality
coefficients are large enough.


\begin{figure}[htp]
\center
\includegraphics[width=0.40\columnwidth]{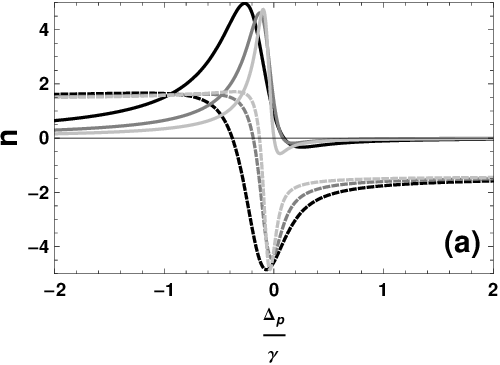}\includegraphics[width=0.40\columnwidth]{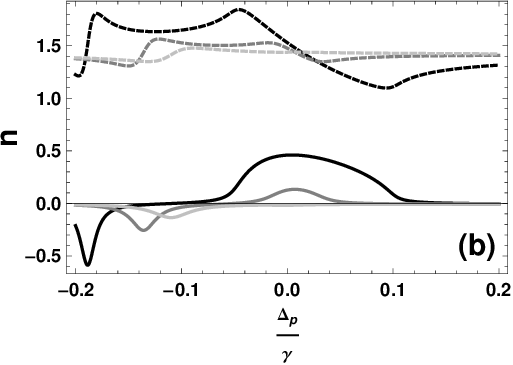}
\caption{Real( solid lines) and imaginary(dashed lines) parts of the
refractive index as a function of the rescaled detuning parameter
$\Delta_{p}/\gamma$.The black,gray and light gray lines correspond
to(a):$\theta = 1/5 \pi$,$\Omega_{c} = 0.4\gamma, 0.8\gamma$,
$1.3\gamma$;(b):$\theta = 3/2 \pi$,$\Omega_{c} = 1\gamma,
1.4\gamma,1.8\gamma$,respectively.The other parameters are same as
in Fig.2.}
\label{Fig.4}
\end{figure}


In Fig.3(a), the refraction index is plotted for $\theta = 1/5
\pi$,$\Omega_{c} = 0.4\gamma, 0.8\gamma$, $1.3\gamma$ ,and the
black,gray and light gray curves correspond to the different values
of $\Omega_{c}$,respectively.The other parameters are identical to
those of Fig.2.As shown in Fig.3(a),the real part of the refractive
index is negative in the region of [0,2$\gamma$],and the imaginary
part depicts the absorption and gain properties. $\Omega_{c}$ =
$1.3\gamma$,the amplitude of the negative refraction index obtains
maximum.We change the phase difference between the control and pump
fields to $\theta = 3/2 \pi$ ,and vary $\Omega_{c}$ by $1\gamma,
1.4\gamma,1.8\gamma$ in Fig.3(b).It's observed that the negative
refraction index amplitude is decreasing with the variation of
$\Omega_{c}$,and the frequency region of negative refraction index
is gradually widening companying with the same variation.The
imaginary part of the refractive index shows that some absorptions
do exist in the same regions as mentioned in [23], due to the effect
of the electromagnetically induced chirality.

\begin{figure}[htp]
\center
\includegraphics[width=0.40\columnwidth]{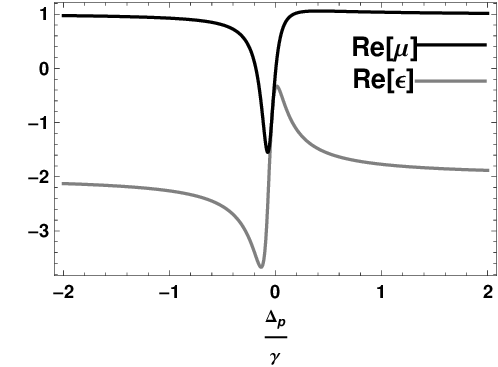}
\includegraphics[width=0.40\columnwidth]{fig4.eps}
\caption{Real(solid lines) and imaginary(dashed lines) parts of the
permittivity and permeability as a function of the rescaled detuning
parameter $\Delta_{p}/\gamma$ for $\theta = 1/5 \pi$,$\Omega_{c} =
1.3 \gamma$,and the other parameters are same as in Fig.2.}
\label{Fig.5}
\end{figure}

We plot the electric permittivity and magnetic permeability in Fig.4
by the separate depiction of their real and imaginary
parts,respectively.We noticed that the electric permittivity and
magnetic permeability are not simultaneously negative in the
frequency region of negative refraction mentioned in Fig.3.The real
part of electric permittivity is negative in the considering
frequency range but the magnetic permeability is not always like
this. And the similar conclusion can be drawn by varying the
corresponding parameters identical to Fig.3.

In experimental investigations,the level configuration shown in
Fig.1 may be realized in the atomic hydrogen or neon because each
has the same level structure as that of Fig.1 [19].Detailed
assessment of our approach in such system will be among our future
investigations. We also expect our technique to be applicable in
other systems including molecules or solid-state structures in the
future.

\section{Conclusion}

In conclusion,we utilized magneto-electric cross coupling to obtain
chirality-induced negative index of refraction without requiring
requiring simultaneously negative both permittivity and
permeability.By tuning the phase and amplitude of the external
fields properly,the two chirality coefficients have the same
amplitude but the opposite phase ,and the atomic system shows that
the negative refraction can be carried out without requiring the
simultaneous presence of an electric-dipole and a magnetic-dipole
transition near the same transition frequency.

\section*{Acknowledgments}

The work is supported by the National Natural Science Foundation of
China ( Grant No.60768001 and No.10464002 ).


\begin{thebibliography}{0}%
\makeatletter
\providecommand \@ifxundefined [1]{%
 \@ifx{#1\undefined}
}%
\providecommand \@ifnum [1]{%
 \ifnum #1\expandafter \@firstoftwo
 \else \expandafter \@secondoftwo
 \fi
}%
\providecommand \@ifx [1]{%
 \ifx #1\expandafter \@firstoftwo
 \else \expandafter \@secondoftwo
 \fi
}%
\providecommand \natexlab [1]{#1}%
\providecommand \enquote  [1]{``#1''}%
\providecommand \bibnamefont  [1]{#1}%
\providecommand \bibfnamefont [1]{#1}%
\providecommand \citenamefont [1]{#1}%
\providecommand \href@noop [0]{\@secondoftwo}%
\providecommand \href [0]{\begingroup \@sanitize@url \@href}%
\providecommand \@href[1]{\@@startlink{#1}\@@href}%
\providecommand \@@href[1]{\endgroup#1\@@endlink}%
\providecommand \@sanitize@url [0]{\catcode `\\12\catcode `\$12\catcode
  `\&12\catcode `\#12\catcode `\^12\catcode `\_12\catcode `\%12\relax}%
\providecommand \@@startlink[1]{}%
\providecommand \@@endlink[0]{}%
\providecommand \url  [0]{\begingroup\@sanitize@url \@url }%
\providecommand \@url [1]{\endgroup\@href {#1}{\urlprefix }}%
\providecommand \urlprefix  [0]{URL }%
\providecommand \Eprint [0]{\href }%
\providecommand \doibase [0]{http://dx.doi.org/}%
\providecommand \selectlanguage [0]{\@gobble}%
\providecommand \bibinfo  [0]{\@secondoftwo}%
\providecommand \bibfield  [0]{\@secondoftwo}%
\providecommand \translation [1]{[#1]}%
\providecommand \BibitemOpen [0]{}%
\providecommand \bibitemStop [0]{}%
\providecommand \bibitemNoStop [0]{.\EOS\space}%
\providecommand \EOS [0]{\spacefactor3000\relax}%
\providecommand \BibitemShut  [1]{\csname bibitem#1\endcsname}%
\let\auto@bib@innerbib\@empty
\end{thebibliography}%


\begin{thebibliography}{99}
\bibitem{01}Veselago V G, {\it Soviet Physics Usp.} {\bf 10}(1968) 509.
\bibitem{02}Shelby R,Smith D R, and Schultz S,{\it Science} {\bf 292} (2001) 77 .
\bibitem{03}Yen T J,Padilla W J,Fang N,Vier D C,Smith D R,Pendry J B,Basov D N, and Zhang X,{\it Science} {\bf 303}(2004)1494 .
\bibitem{04}Veselago V G and Narimanov E E {\it Nature Mater.} {\bf 5}(2006) 759.
\bibitem{05}Chen Y Y,Huang Z M,Shi J L,Li C F and Wang Q, {\it Chin. Phys.} {\bf 16} (2007) 173 .
\bibitem{06}Zhao S C,Liu Z D,{\it Int.J.Quan.Inf.} {\bf 7}(2009)747.
\bibitem{07}Lin Z L,Ding J C and Zhang P, {\bf Chin. Phys.B} {\bf 18}(2008)954.
\bibitem{08}Pendry J B,{\it Phys.Rev.Lett.} {\bf 85} (2000)3966.
\bibitem{09}Aydin K,Bulu I and Ozbay E {\it Appl.Phys.Lett.} {\bf 90}(2007)254102
\bibitem{10}Chen L,He S and Shen L {\it Phys.Rev.Lett.} {\bf 92}(2004)107404
\bibitem{11}Jiang Y Y,Shi H Y,Zhang Y Q,Hou C F and Sun X D, {\bf Chin. Phys.} {\bf 16} (2007)1959.
\bibitem{12}Dong Z G,Zhu S N and Liu H, {\bf Chin. Phys.} {\bf 17} (2006)1772.
\bibitem{13}Shelby R A,Smith D R,Schultz S,{\it Science} {\bf 292} (2001) 77-79.
\bibitem{14}Pendry J B,{\it Nature} {\bf 423} (2003) 22-23.
\bibitem{15}Cubukcu E, {\it Nature} {\bf 423} (2003)604-605.
\bibitem{16}Eleftheriades G V,Iyer A K,Kremer P C,{\it IEEE Trans.Microwave Theory Tech.} {\bf 50}(2002)2702-2712.
\bibitem{17}Pendry J B,{\it Science} {\bf 306} (2004) 1353-1355.
\bibitem{18}Yannopapas V,{\it J. Phys.: Condens. Matter} {\bf 18}(2006) 6883-6890
\bibitem{19}Thommen Q,Mandel P,{\it Phys.Rev.Lett.} {\bf 96}(2006) 053601.
\bibitem{20}Oktel M $\ddot{o}$,M$\ddot{u}$tecapl$\check{g}$u $\ddot{o}$ E,{\it Phys.Rev.A} {\bf 70}(2004)053806.
\bibitem{21}Shen J Q, {\it Phys. Lett. A} {\bf 357} (2006) 54.
\bibitem{22}Monzon C and Forester D W,{\it Phys.Rev.Lett.} {\bf 95},(2005)123904 .
\bibitem{23}K$\ddot{a}$astel J,Fleischhauer M,Yelin S F,Walsworth R L, {\it Phys.Rev.Lett.} {\bf 99} (2007) 073602
\bibitem{24}K$\ddot{a}$astel J,Fleischhauer M,Walsworth R L,{\it Phys.Rev.A} {\bf 79}(2009)063818
\bibitem{25}Zhao S C, Liu Z D, and Wu Q X, {\it Chin. Phys.B } {\bf 19},(2010)014211 .
\bibitem{26}Shen J Q, {\it Chin. Phys.} {\bf 16},(2007)1976.
\bibitem{27}Cook D M,1975 {\it The Theory of the Electromagnetic Field}(Prentice-Hall,New Jersey){\bf chapter 11}.
\bibitem{28}Jackson J D,1999 {\it Classical Electrodynamics(3rd edn)}(New York: Wiley){\bf p160}
\end{thebibliography}
\end{document}